# Perfect photon absorption in nonlinear regime of cavity quantum electrodynamics

G. S. Agarwal[1], Ke Di[2], Liyong Wang[2-5], and Yifu Zhu[2]

[1]Department of Physics, Oklahoma State University, Stillwater, Oklahoma 74078, USA
[2]Department of Physics, Florida International University, Miami, Florida 33199, USA
[3] Center for cold atom physics, Chinese Academy of Sciences, Wuhan, 430071, China
[4]Wuhan Institute of Physics and Mathematics, Chinese Academy of Sciences, Wuhan, 430071, China
[5]University of Chinese Academy of Sciences, Beijing 100049, China 230026, China

It has been shown that perfect photon absorption can occur in the linear excitation regime of cavity quantum electrodynamics (CQED), in which photons from two identical light fields coupled into two ends of the cavity are completely absorbed and result in excitation of the polariton state of the CQED system. The output light from the cavity is totally suppressed by the destructive interference and the polariton state can only decay incoherently back to the ground state. Here we analyze the perfect photon absorption and onset of optical bistability in the nonlinear regime of the CQED and show that the perfect photon absorption persists in the nonlinear regime of the CQED below the threshold of the optical bistability. Therefore the perfect photon absorption is a phenomenon that can be observed in both linear and nonlinear regimes of CQED. Furthermore, our study reveals for the first time that the optical bistability is influenced by the input-light interference and can be manipulated by varying the relative phase of the two input fields.

I. INTRODUCTION

Control of the photon absorption, scattering, and localization in an absorbing medium is a current research topic and is important for a variety of fundamental studies and practical applications [1-5]. Recent studies show that coherent perfect photon absorption can be realized in linear dielectrics confined in a Fabry-Perot structure [6-11]. In such solid-state Fabry-Perot devices, all input lights are coherently absorbed by the linear absorber and there is no output light from the etalon devices. The physics behind the perfect photon absorption is the interference between the transmitted field and the input field. Specifically, it requires that the imaginary (absorptive) part of the refractive index of the absorber is equal in absolute value to that of the gain medium at the lasing threshold and the incident light fields match the time-reversed lasing fields. Such perfect coherent absorber and the resulting interferometric control may have practical applications in optics communications and photonic devices such as transducers, modulators, and optical switches and transistors [12-15].

Recently, the coherent perfect photon absorption with quantum fields using path entanglement was theoretically proposed by Huang and Agarwal [16] and experimentally verified by Roger et al [17]. Also similar coherent perfect photon absorption has been studied in a cavity opto-mechanical system [18] and a cavity quantum electrodynamics (CQED) system [19]. A basic CQED system can act as a coherent perfect absorber so long as the CQED system is in the strong coupling regime and the two input fields are tuned near the atomic resonance. Specifically, in the strong photon-atom coupling regime of CQED, the coherently coupled photons and atoms form distinct polariton states (the eigenstates of the CQED system). When the cavity frequency and the input light frequency mutually satisfy a simple relationship, the input lights are completely absorbed. Although the intra-cavity light field is at or near the peak value, no light can be transmitted through the cavity. The absorbed photon energy can only be dumped through the spontaneous decay of the atomic excitation of the polariton state while the photon decay through the cavity is inhibited.

Previous studies of perfect photon absorption are in the linear absorption regime, in which either a linear absorber is considered (such as in a solid-state Fabry-Perot etalon [6-12]) or in CQED, only the linear excitation of the first-order polariton states [19-25] are considered [19] and the multiphoton excitation of the higher-order polariton states in CQED are excluded [26-27]. Naturally this raises a question on whether coherent perfect photon absorption can occur in an optical system with a nonlinear absorber. A recent study explored the manifestation of the dispersive Kerr nonlinearity on the coherent photon absorption and scattering in a layered dielectric medium [28]. Here we analyze the nonlinear excitation of CQED and show that perfect photon absorption in CQED can persist in the nonlinear regime of CQED before onset of optical bistability. We characterize the phase dependence of the perfect photon absorption and show the interference effects on the optical bistability in the CQED system. Our results indicate that the perfect photon absorption is a general phenomenon in the strong coupling regime of CQED and occurs in both the linear and nonlinear excitation regimes of CQED. Our analysis involves all orders of nonlinearities in the CQED system and the prefect photon absorption can be tuned from the linear regime into the nonlinear regime by varying the frequency, intensity, and relative phase of the input light fields.

II. THEORETICAL MODEL AND ANALYSIS

Fig. 1 shows the schematic diagram for a coupled CQED system that consists of N two-level atoms confined in a single mode cavity and is excited by two input light fields $a_{in}^r$ and $a_{in}^l$ from two opposite ends of the cavity. The cavity mode couples the atomic transition $|g\rangle - |e\rangle$ with frequency detuning $\Delta_c = \nu_c - \nu_{ge}$. The two input fields have the same frequency $\nu_p$ and are tuned from the atomic transition by $\Delta_p = \nu_p - \nu_{ge}$. We define the collective atomic operators $S^z = \frac{1}{2}\sum_{i=1}^{N}(\sigma_{ee}^i - \sigma_{gg}^i)$, $S^+ = \sum_{i=1}^{N}\sigma_{eg}^i$ ($\hat{\sigma}_{lm}^{(i)} = |l\rangle\langle m|^{(i)}$ (l,m=e or g) is the atomic operator for the ith atom). The Hamiltonian for the CQED system is

$$H = (\Delta_c - \Delta_p)a^+ a + \Delta_p S_z + \{gaS^+ + ia^+(\sqrt{2\kappa_r}a_{in}^r + \sqrt{2\kappa_l}a_{in}^l) + H.C.\}. \quad (1)$$

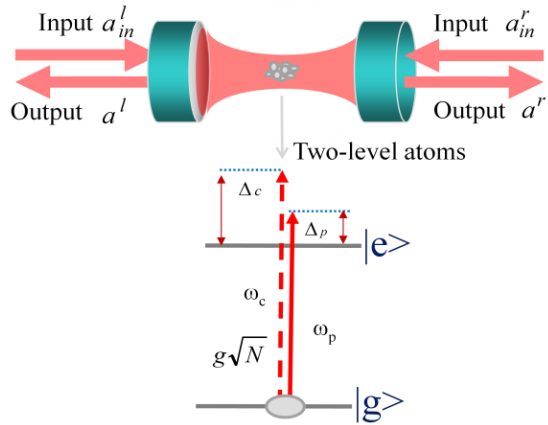

Fig. 1 A CQED system with N two-level atoms confined in the cavity mode and coupled by two input light fields from opposite ends of the cavity. The dashed arrow denotes the interaction of the cavity mode and the solid arrow denotes the interaction of the input light fields.

Here $\hat{a}$ ($\hat{a}^+$) is the annihilation (creation) operator of the cavity photons, $a_{in}^l$ and $a_{in}^r$ are two input fields to the cavity (see Fig. 1), $\kappa_i = \frac{T_i}{\tau}$ (i=r or l) is the loss rate of the cavity field on the mirror i ($T_i$ is the mirror transmission and $\tau$ is the photon round trip time inside the cavity), and $g = \mu_{eg}\sqrt{\omega_c/2\hbar\varepsilon_0 V}$ is the cavity-atom coupling coefficients and is assume to be uniform for the N identical atoms inside the cavity. The equations of motion for the coupled CQED system are given by $\frac{d\hat{\sigma}}{dt} = -i[H,\hat{\sigma}] + \hat{L}\hat{\sigma}$. We drop the quantum fluctuation terms and derive the equations of motion of the CQED system for the expectation values of $S^+$, $S^-$, $S^z$, and $a$,

$$\dot{S}^z = -\Gamma(S_z + N/2) - i(gaS^+ - ga^+ S^-), \quad (2\text{-}1)$$

$$\dot{S}^+ = -(\Gamma/2 - i\Delta_p)S^+ - 2iga^+ S^z, \quad (2\text{-}2)$$

$$\dot{a} = -((\kappa_l + \kappa_r)/2 + i(\Delta_c - \Delta_p))a - igS^- + \sqrt{\kappa_1/\tau}a_{in}^r + \sqrt{\kappa_2/\tau}a_{in}^l. \quad (2\text{-}3)$$

Here $\Gamma$ is the decay rate of the excited state $|e\rangle$. Assume a symmetric cavity for the subsequent analysis, $T_l = T_r = T$, then $\kappa_l = \kappa_r = \kappa$. The steady-state solution of the intra-cavity light field is then given by

$$a = \frac{\sqrt{\kappa/\tau}(a_{in}^r + a_{in}^l)}{\kappa - i(\Delta_p - \Delta_c) + \frac{g^2 N}{(\frac{\Gamma}{2} - i\Delta_p)(1 + \frac{2g^2|a|^2}{\frac{\Gamma^2}{4} + \Delta_p^2})}}, \quad (3)$$

The denominator in the right side of Eq. (3) contains the intra-cavity intensity term $|a|^2$ and shows the nonlinear dependence of the intra-cavity field on the input light fields. As in free space, one can define the intra-cavity saturation intensity as $|a_s|^2 = \frac{\Gamma^2}{4g^2}$ (corresponding to the Rabi frequency of the intra-cavity field, $ga$, equals to one half of the atomic decay linewidth). When the intra-cavity light intensity $|a|^2 \geq |a_s|^2$, the CQED system is driven into the nonlinear excitation regime.

The output fields from the left mirror and right mirror of the cavity are related to the intra-cavity field and the input fields and are given by

$$a^l = \sqrt{T}a - a^l_{in} \quad (4)$$

$$a^r = \sqrt{T}a - a^r_{in} \quad (5)$$

Assume $a^r_{in} = |a^r_{in}|$ and $a^l_{in} = \alpha |a^r_{in}| e^{i\Phi}$ ($\alpha$ is a real number), the equation relates the input intensity $|a^l_{in}|^2 = |a^r_{in}|^2 = |a_{in}|^2$ and the intra-cavity intensity $|a|^2$ is given by

$$|a^r_{in}|^2 = \frac{T|a|^2}{\kappa^2(1+\alpha^2+2\alpha\cos(\Phi))}\{\kappa^2 + (\Delta_p - \Delta_c)^2 + \frac{g^2 N[\kappa\Gamma - 2\Delta_p(\Delta_p - \Delta_c)]}{\frac{\Gamma^2}{4}+\Delta_p^2+2g^2|a|^2} + \frac{g^4 N^2(\frac{\Gamma^2}{4}+\Delta_p^2)}{(\frac{\Gamma^2}{4}+\Delta_p^2+2g^2|a|^2)^2}\}. \quad (6)$$

Then the output intensity from the left mirror is

$$|a^l|^2 = \left|\frac{(1+\alpha e^{i\Phi})\kappa}{\kappa - i(\Delta_p - \Delta_c) + \frac{g^2 N}{(\frac{\Gamma}{2}-i\Delta_p)(1+\frac{2g^2|a|^2}{\frac{\Gamma^2}{4}+\Delta_p^2})}} - \alpha e^{i\Phi}\right|^2 |a_{in}|^2, \quad (7)$$

and the output light intensity from the right mirror is

$$|a^r|^2 = \left|\frac{(1+\alpha e^{i\Phi})\kappa}{\kappa - i(\Delta_p - \Delta_c) + \frac{g^2 N}{(\frac{\Gamma}{2}-i\Delta_p)(1+\frac{2g^2|a|^2}{\frac{\Gamma^2}{4}+\Delta_p^2})}} - 1\right|^2 |a_{in}|^2 \quad (8)$$

Here the light intensity is equal to the expectation value of the number of photons, i.e., $|a|^2 = n$, $|a_{in}|^2 = n_{in}$, $|a^l|^2 = n^l$, and $|a^r|^2 = n^r$. Eq. (7) and (8) show the nonlinear dependence of the output intensity versus the input intensity and is similar to the standard nonlinear equations for the optical bistability with a single input field [29-30]. Therefore, it is expected that when the input intensity is sufficiently large, the CQED system with two input fields will be driven into the bistable domain. It will be shown that unlike the optical bistability with a single input field, the interference of two input fields results in the phase dependence of the optical bistability, which enables control of the optical bistability by varying the phase Φ. In the following, we will analyze the conditions of coherent perfect photon absorption in the nonlinear coupling regime of CQED, and characterize it below and above the onset of the optical bistability. For the CQED system acting as a perfect photon absorber, the two input fields must be identical, $a^r_{in} = a^l_{in}$ ($\alpha=1$ and $\varphi=0$) and the two output fields are zero. With $a^r = a^l = 0$ in Eq. (4) and (5), one derives two specific conditions for the perfect photon absorption in CQED:

$$\Delta_c = \frac{\Gamma - 2\kappa}{\Gamma}\Delta_p, \quad (9)$$

and

$$\frac{\kappa\Gamma}{2}+\frac{2\kappa}{\Gamma}\Delta_p^2 = \frac{g^2 N}{1+\dfrac{2g^2|a|^2}{\dfrac{\Gamma^2}{4}+\Delta_p^2}}. \qquad (10)$$

The first condition (9) specifies the required frequency matching of the cavity and the input fields and the 2nd condition specifies the nonlinear relation between the intra-cavity field intensity (the input light intensity) and the CQED system parameters. When Eq. (8) and Eq. (10) are valid, there is no output light from the cavity, but the intra-cavity light intensity $|a|^2$ is nonzero and the corresponding input-light intensity is

$$|a_{in}|^2 = T|a|^2 = T\left(\frac{N(\frac{\Gamma^2}{4}+\Delta_p^2)}{\kappa\Gamma+\frac{4\kappa}{\Gamma}\Delta_p^2}-\frac{\frac{\Gamma^2}{4}+\Delta_p^2}{2g^2}\right). \qquad (11)$$

This indicates that the perfect photon absorption in a CQED system can be achieved by tuning the input light intensity $|a_{in}|^2$ to match Eq. (11), showing the capability of the nonlinear control of the perfect photon absorption in CQED by the input light intensity. For nontrivial solutions, $|a_{in}|^2 > 0$, which leads to $g^2 N > \frac{\kappa\Gamma}{2}+\frac{2\kappa\Delta_p^2}{\Gamma}$. The minimum $g^2 N$ value occurs at $\Delta_p=0$, which leads to $g^2 N > \frac{\kappa\Gamma}{2}$, exactly the condition for the collective strong coupling regime in CQED. Hence, if a CQED system is in the strong coupling regime, the prefect photon absorption can be realized by varying the system parameters to match Eq. (9) and Eq. (10). In certain applications, it is desirable to realize the perfect photon absorption with a broad bandwidth [31]. For the CQED system, one can have a cavity with $\kappa \gg \Gamma$ while operating in the linear excitation regime of the CQED system with $|a|^2 \ll \frac{\Gamma^2}{4g^2}$ and at resonance ($\Delta_p=\Delta_c=0$); then by matching the condition $g^2 N = \frac{\kappa\Gamma}{2}$, the perfect photon absorption can be realized in a frequency bandwidth much greater than the atomic linewidth $\Gamma$.

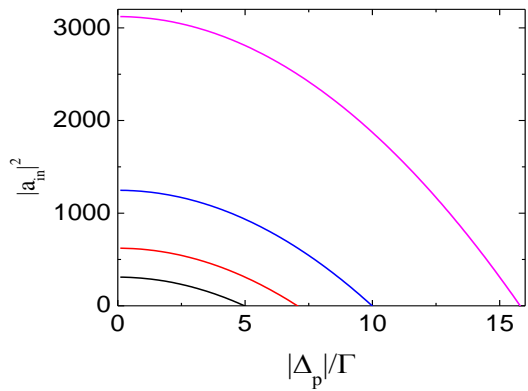

Fig. 2 Under the perfect-photon-absorption condition, the input light intensity $|a_{in}|^2$ versus the input frequency detuning $\Delta_p/\Gamma$. $\kappa=0.2\Gamma$ and $\Delta_c=0.6\Delta_p$ for the purple line; $\kappa=0.5\Gamma$ and $\Delta_c=0$ for the blue line; $\kappa=\Gamma$ and $\Delta_c=-\Delta_p$ for the red line; and $\kappa=2\Gamma$ and $\Delta_c=-3\Delta_p$ for the black line. The other parameters are

$g\sqrt{N} = 10\Gamma$, g=0.02Γ, and T=0.01.

Fig. 2 plots the input light intensity $|a_{in}|^2$ under the condition of the prefect photon absorption versus the input frequency detuning $|\Delta_p|/\Gamma$. The system parameters are chosen to be consistent with that readily achievable experimentally [27,32]. It shows that the perfect photon absorption occurs only in the frequency range of the input light fields $\Delta_p^2 \leq \frac{g^2 N\Gamma}{2\kappa} - \frac{\Gamma^2}{4}$. When the CQED system satisfies the perfect photon absorption condition, the intra-cavity field $a$ is related to the input field $a_{in}$ by the simple equation $a = \frac{a_{in}}{\sqrt{T}}$. So the CQED system with the perfect photon absorption is in the nonlinear excitation regime if the input field intensity $|a_{in}|^2 \geq \frac{T\Gamma^2}{4g^2}$. Fig. 2 shows that as the frequency detuning of the input light decreases from the threshold value $\Delta_p^T = \sqrt{\frac{g^2 N\Gamma}{2\kappa} - \frac{\Gamma^2}{4}}$, the input light intensity $|a_{in}|^2$ increases from zero and reaches value $|a_{in}|^2 \geq \frac{T\Gamma^2}{4g^2}$; correspondingly, the CQED system evolves from the linear excitation regime into the nonlinear excitation regime. As a numerical example with the system parameters provided in Fig. 2, when the input light intensity $|a_{in}|^2 \geq 6.25$, the CQED system is in the nonlinear regime. Therefore, the perfect photon absorption can be observed in the highly nonlinear coupling regime of the CQED. Fig. 2 also shows that the smaller the cavity decay κ is, (corresponding to a higher cavity finesse), the wider the frequency range of the input light in which the perfect photon absorption can be observed, and also the greater the input light intensity can be, which leads to a stronger nonlinear excitation regime of CQED.

### III. INTERFERENCE CONTROL OF PERFECT PHOTON ABSORPTION AND OPTICAL BISTABILITY

It is well known that in a CQED system (with a specific set of system parameters g, $g\sqrt{N}$, γ, κ and T), as the input light intensity increases, the nonlinear interaction of the atoms and the intra-cavity light field becomes dominant, and the CQED system is driven into the bistable domain when the input light intensity is above the threshold for the onset of the optical bistability. Although Fig. 2 shows that there is a large dynamic range of the input light intensity in which the perfect photon absorption occurs while the CQED system is in the nonlinear regime, it is necessary to find out whether the perfect photon absorption occurs in the bistable domain of the CQED or the normal working domain of the CQED. When the CQED system is in the bistable domain, there are three real solutions of the output light intensity $|a_{out}|^2$ for a given input light intensity $|a_{in}|^2$. The threshold value of the input light intensity $|a_{in}|^2$ for the onset of the optical bistability can be found from the nonlinear equation (7) or (8) with two identical input fields ($a_{in}^r = a_{in}^l = a_{in}$). However, it is difficult to derive the analytical solution, we therefore opt to provide numerical calculations that characterize the perfect photon absorption in the normal and bistable operating domains of the CQED.

Fig. 3 plots the output light intensity versus the input light intensity for a series of different system parameters. The perfect photon absorption is manifested by the interference of the two input fields and occurs at a given input intensity satisfying Eq. (11) and with Φ=2nπ (the two input fields are in phase and the two output fields are equal as shown by the blue curves in Fig. 3). When Φ=(2n+1)π, the two input fields interfere destructively (no light can be coupled into the cavity), and the two output intensities $|a^l|^2$ and $|a^r|^2$ are identical and equal to the input intensity $|a_{in}|^2$ as shown by the black curves in Fig. 3. At other $\Phi$ values, the two output intensities $|a^l|^2$ and $|a^r|^2$ are generally different and the perfect photon absorption does not occur (see the red and green curves for $\Phi=\pi/2$ in Fig. 3). Fig. 3 shows that for a given set of the system parameters, when the frequency detuning of the input light field is greater than a threshold value $(\Delta_p)_T$, the CQED system is in the normal operating domain (the nonlinear equation (7) ((8)) has only one real-value solution for $|a^l|^2$ and $|a^r|^2$, and there is a definite value of the input light intensity at which the perfect photon absorption can be observed with $\Phi=0$ (the blue curves in Fig. 3). In the normal operating domain of the CQED, the perfect photon absorption can be observed in the linear excitation regime with weak input light intensities at the frequency detuning $|\Delta_p|$ close to the threshold value $\Delta_p^T = \sqrt{\frac{g^2 N\Gamma}{2\kappa} - \frac{\Gamma^2}{4}}$ as shown in Fig. 3(a); as $|\Delta_p|$ decreases from $\Delta_p^T$, the CQED system is driven into the nonlinear excitation regime and the perfect photon absorption is observed at the input light intensity above the saturation level (Fig. 3(b); at even smaller $\Delta_p$ values, the CQED is driven into the bistable operation domain when the input intensity is above the threshold value for the optical bistability (the nonlinear equation (7) has three real value solution for $|a_{out}|^2$) as shown in Fig. 3(c) and 3(d). When the CQED system is in the bistable domain, the point at which the output intensity is zero (the perfect photon absorption) is near the turning point of the input light intensity on the return hysteresis curve (see Fig. 3(c) and 3(d)). The calculations are also done with the cavity decay rate $\kappa>2\Gamma$ and $\kappa<2\Gamma$ while keeping Eq.

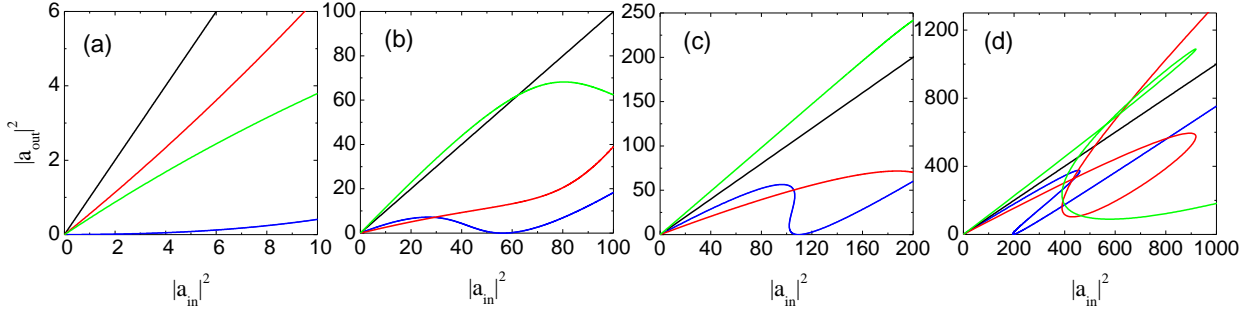

Fig. 3 Output light intensity $|a_{out}|^2$ ( $|a^r|^2$ or $|a^l|^2$ ) versus input light intensity $|a_{in}|^2$ with two input fields of the same amplitude $|a_{in}^l|^2=|a_{in}^r|^2=|a_{in}|^2$ and varying relative phase $\Phi$ ($\Phi=\pi$ for black curves with which the two output intensities are equal, $|a^r|^2=|a^r|^2$; $\Phi=\pi/2$ for red and green curves with which the two output intensities are different with red curves for the right output $|a^r|^2$ and green curves for the left output $|a^l|^2$; $\Phi=0$ for blue curves with which the two output intensities are equal, $|a^l|^2=|a^r|^2$.) . The perfect photon absorption requires $\Phi=2n\pi$ (n=0, 1, 2...., is an integer). (a) $\Delta_p=5\Gamma$ and $\Delta_c=-15\Delta_p$; (b) $\Delta_p=4.5\Gamma$ and $\Delta_c=-13.5\Delta_p$; (c) $\Delta_p=4\Gamma$ and $\Delta_c=-12\Delta_p$; and (d) $\Delta_p=3\Gamma$ and $\Delta_c=-9\Delta_p$. The other parameters are $\kappa=2\Gamma$, g=0.02$\Gamma$, T=0.01, $g\sqrt{N}=10\Gamma$ .

(9) always valid and show similar behavior of the output light intensity versus the input light intensity as shown in Fig. 3.

It is worth noting that Fig. 3 also reveals the phase dependence of the optical bistability in the CQED system with two input light fields and demonstrates the possibility of active control of the optical bistability by the light interference. For example, Fig. 3(c) and 3(d) show that the optical bistability occurs when the relative phase $\Phi=0$; by tuning the phase $\Phi$ to a different value (for example $\Phi=\pi/2$, the red and green curves), the CQED system can be out of the bistable domain and operates in the monostable domain with two output fields having different intensities. Furthermore, Fig. 3(d) shows that in the deep optical bistable domain, the phase $\Phi$ can be used to manipulate the bistability and obtain complicated bistable patterns. A detailed study on the interference control of optical bistability is planned and will be carried out in the near future.

Eq. (7) and Eq. (8) show that the perfect photon absorption in CQED can be controlled by varying the relative phase $\Phi$ between the two input fields $a_{in}^r = |a_{in}^r|$ and $a_{in}^l = |a_{in}^r|e^{i\Phi}$. A detailed phase dependence of the perfect photon absorption in the CQED system is shown in Fig. 4 that plots the two output intensities of the CQED system versus the phase $\Phi$ with the perfect photon absorption conditions (7) and (8) satisfied and the input light intensity $|a_{in}^r|=|a_{in}|^2$ below the threshold for the onset of the optical bistablity. It shows that when the two input fields are in phase ($\Phi=0$), there is the perfect photon absorption and the two output fields are zero; when the two input fields are out of phase ($\Phi=\pi$), the two output fields are near the maximum and have the same amplitude. With lower input light intensities (Fig. 4(a)), the CQED system is not in the strong nonlinear excitation regime and the phase dependence exhibits a near sinusoidal curve; at higher input light intensities (Fig. 4(b)), the CQED system is driven into the strong nonlinear excitation regime but still in the mono-stable domain (there is only one real solution from Eq. (7) and eq.(8)), and the nonlinearity causes the phase dependence to deviate appreciably from the sinusoidal behavior. When the CQED system is in the bistable domain, there are three real solutions from Eq. (7) and Eq. (8), complicated phase dependent patterns appear and the CQED system may be driven in or out of the bistable domain (see Fig. 3(c)) at different $\Phi$ values or follow different bistable patterns (see Fig. 3(d)).

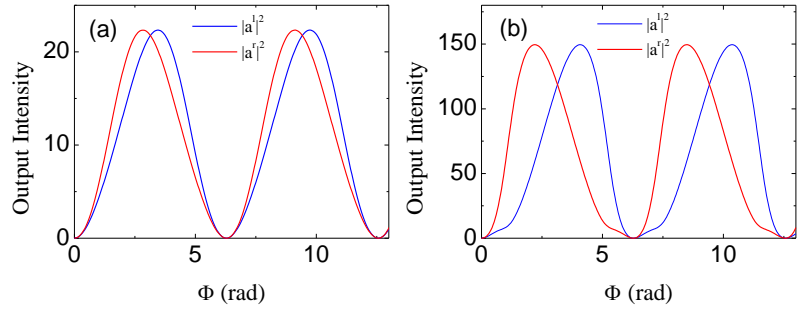

Fig. 4 With the input light intensity below the threshold of the bistable domain, the left output intensity $|a^l|^2$ (red lines) and the right output intensity $|a^r|^2$ (blue lines) versus the phase difference $\Phi$ between the two input fields $a_{in}^r = |a_{in}^r|$ and $a_{in}^l = |a_{in}^r|e^{i\Phi}$. In (a), $|a_{in}^r|=|a_{in}|^2 = 21.75$ and $\Delta_p=9.9\Gamma$; in (b) $|a_{in}^r|=|a_{in}|^2=118.7$ and $\Delta_p=9.5\Gamma$. The other parameters for (a) and (b) are the same with $\kappa=0.5\Gamma$, $g=0.02\Gamma$, $g\sqrt{N}=10\Gamma$, T=0.01, and $\Delta_c=0$.

Similar behavior is also observed for the calculated phase dependence of the output light intensities with other different system parameters.

We note that the CQED parameters used in the calculations of Fig. 2, Fig. 3, and Fig. 4 are taken from reported experimental work done on a multiatom CQED system consisting of a moderate-finesse cavity and cold Rb atoms [27, 32]. With the high density of cold atoms and negligible Doppler shift, a CQED system with cold atoms is well suited for experimental studies of the perfect photon absorption and its interference manipulation in the linear and nonlinear regimes of the CQED, the interplay of the perfect photon absorption with the optical bistability, and the interference control of the CQED bistability reported here.

## IV. CONCLUSION

In conclusion, we have analyzed the light input-output relationship in a multiatom CQED system coupled by two laser fields from two output ends of the cavity in the linear and nonlinear excitation regimes. Our results show that the perfect photon absorption is a general phenomenon and can be observed in the linear excitation or nonlinear excitation regimes of the CQED when the CQED system satisfies the strong coupling condition. We derived the specific conditions for the validity of the prefect photon absorption and characterized its behavior in the monostable operation and bistable operation domains of the CQED. Our analysis also demonstrates the phase dependence of the optical bistability in a CQED system coupled by two input fields and thus provides a practical method to control the optical bistability by light interference, which may be useful for all optical switching and optical multiplexing applications.

It will be interesting to study the perfect photon absorption in the quantum regime of the nonlinear CQED. It has been shown that the quantum interference can be used to suppress the linear absorption and enhance the quantum nonlinear absorption in a CQED system [27], which enables studies of the pure nonlinear CQED phenomena free of the linear absorption effects. Thus it will be interesting to see if the perfect photon absorption can occur under the pure quantum nonlinear excitation of the CQED.

ACKNOWLEDGEMENT

Y. Zhu acknowledges support from the National Science Foundation under Grant No. 1205565.